\documentclass[12pt]{article}

\usepackage{epsfig}
\def\be{\begin{equation}}
\def\ee{\end{equation}}
\def\ben{\begin{displaymath}}
\def\een{\end{displaymath}}
\def\ba{\begin{array}{c}}
\def\ea{\end{array}}

\newcommand{\kt}{\rangle}
\newcommand{\br}{\langle}
\newcommand{\ed}{\end{document}}
\newcommand{\bbr}{\br\!\br}
\newcommand{\kkt}{\kt\!\kt}

\newcommand{\pbr}{\prec\!}
\newcommand{\pkt}{\!\succ\,}

\begin{document}

\begin{center}

{\Large \bf

Identification of observables
 in quantum toboggans.

  }

\vspace{9mm}

{Miloslav Znojil}

\vspace{9mm}

{\'{U}stav jadern\'e fyziky AV \v{C}R, 250 68 \v{R}e\v{z}, Czech
Republic}

{znojil@ujf.cas.cz}


\end{center}


\section*{Abstract}

Schr\"{o}dinger equations for ``quantum toboggans" with real
energies are given the generalized eigenvalue-problem form $H\psi
= EW\psi$ where $H \neq H^\dagger$ and where $W \neq W^\dagger
\neq I$. The consistent probabilistic interpretation of these
models is provided. The new double-series formula for the
necessary {\it ad hoc} metric $\Theta = \Theta(H,W)$ is derived
which defines the acceptable inner products in the physical
Hilbert space of states. The formula degenerates to the usual
single series in the non-tobogganic trivial-weight limit $W\to I$.

\newpage

\section{Introduction}

For all the sufficiently elementary quantum models which are based
on a Hamiltonian $H$ which is self-adjoint, say, in one of the
most common Hilbert spaces $I\!\!L^2(I\!\!R^d)$ of complex
integrable functions of $d$ real variables (= coordinates) the
identification of all the other eligible candidates for the
operators of physical observables is trivial. From the formal
point of view, any self-adjoint operator ${\cal O} = {\cal
O}^\dagger$ is acceptable \cite{Messiah}.

Fifteen years ago Scholtz et al \cite{Geyer} emphasized that the
same principle applies also in all the less elementary models in
Quantum Theory. They paid attention to the Interacting Boson Model
in nuclear physics where the complicated structure of the explicit
representation ${\cal H}^{(physical)}$ of the Hilbert space of
states of a nucleus makes this space unsuitable for explicit
calculations. The necessary simplification of some (typically,
variational) calculations has been achieved, in this context, via
the Dyson's mapping $\Omega^{(Dyson)}$ which interrelates the
fermionic Hilbert space ${\cal H}^{(physical)}$ and a perceivably
simpler and much more ``user-friendly" auxiliary bosonic Hilbert
space ${\cal H}^{(auxiliary)}$ (cf. also a few further, more
technical remarks relocated to Appendix A below).

Beyond the framework of nuclear physics the concept of the models
which appear to be non-Hermitian ``in a wrong space" occurs also,
from time to time, in the framework of field theory \cite{Nagy}.
In the late nineties, the really dramatic development of the
models of this type was inspired by the publication of a few
papers by Bender et al \cite{BM,BB}. These authors analyzed
several non-Hermitian operators $H$ with real spectra and
conjectured that these models could be perfectly acceptable in
physics, for phenomenological purposes at least (for more details
cf. the recent review \cite{Carl}).

In what follows we intend to re-analyze the question of the
identification of the operators of observables in the latter
non-Hermitian context. In an introductory section \ref{nonu} we
review briefly the class of models of our present interest
(``quantum toboggans") and we offer there also a brief review of
their mathematical origin (cf. subsection \ref{wsvnonu}). This
will be complemented by a simple illustrative example (subsection
\ref{qutobo}) and by a few more technical remarks on the
differences between some related Hilbert spaces (Appendix A), on
the meaning of complex coordinates (Appendix B) and on the
facilitating mathematical role of the so called ${\cal
PT}-$symmetry of the Hamiltonians (cf. Appendix C).

The presentation of our new results will be separated into two
parts. In the first part (section \ref{4.}) we shall start from an
illustrative tobogganic Schr\"{o}dinger equation. We shall
emphasize that the related complex and topologically nontrivial
tobogganic paths of coordinates $z^{(N)}(x) \in l\!\!C$ are
characterized by a winding number $N$ and parametrized by a real
$x \in I\!\!R$. The rectification transformation is then
recommended which maps the original tobogganic, multisheeted
curves of the coordinates in the interior of a single complex
plane with a cut. We emphasize that in the language of functional
analysis, the new form of our tobogganic Schr\"{o}dinger equation
reads $H\psi=EW\psi$ and that it is well defined merely in an
auxiliary, unphysical Hilbert space ${\cal H}^{(auxiliary)}$. In
subsections \ref{4.1} and \ref{biolo} we draw some consequences
from the fact that $W \neq I$.

The second and main part of our message is formulated in section
\ref{manolo}. We emphasize there that whenever $W \neq I$, the
Schr\"{o}dinger's generalized eigenvalue problem $H\psi=EW\psi$
requires a new approach when its physical interpretation is
concerned. The mathematical core of our innovative proposed
strategy is explained in subsection \ref{manoloja} which, in
essence, updates the $W=I$ formalism as summarized in Appendix A.
In a complete parallel we introduce certain set of appropriate
Fourier-like but non-unitary $\Omega-$transformations. Some of the
main consequences are clarified in subsection \ref{4.2}. This
makes us prepared to address, finally, the principal question of
the explicit construction of the physical metric
$\Theta=\Theta(H,W)$ in terms of the solutions of our
Schr\"{o}dinger equations (cf. subsection \ref{quito}).

A concise summary of our results will be presented in section
\ref{sum} where we review, once more, the basic philosophy of the
whole approach and where we emphasize that the underlying
mathematics is based on a non-unitary generalization of the usual
switch between the $x-$ and $p-$representations in Quantum
Mechanics. We stress that our recipe coincides with its simpler
$W=I$ version in the limit $W \to I$.

\section{Tobogganic phenomenological models \label{nonu} }

\subsection{Spiked oscillators \label{wsvnonu} }

The Bender's and Boettcher's ${\cal PT}-$symmetric oscillator
$V^{(BB)}(x) = x^2+{\rm i}x$ (cf. Appendix C) is easily perceived
as a model where the coordinate is complex,
 \be
 r = r(x)=x-{\rm i}\varepsilon\,,
 \ \ \ \ \ \
 \ \ \ \ \ x
 \in (-\infty,\infty)\,.
 \label{line}
 \ee
Thus, one can write $V^{(BB)}(x) = r^2(x)+const$ at a suitable
$\varepsilon> 0$. Of course, the model is exactly solvable. Still,
several less trivial models have been considered in the recent
literature (cf. its review in \cite{Carl}). Many of them were
characterized by the presence of the centrifugal term in the
potential (cf. Appendix B below). For illustrative purposes let us
choose here, therefore, one of the most popular one-dimensional
anharmonic-oscillator  Schr\"{o}dinger equations which is defined
along the {\em complex} straight line (\ref{line}),
 \be
 \left [-\frac{d^2}{dx^2}
 +\frac{\ell(\ell+1)}{r^2}+\omega^2\,r^2+{\rm
 i}r^3\right ]\,\varphi_n(r)=E_n(\omega)\,\varphi_n(r)
 \,, \ \ \ \ \ \ \ r=r(x)\,.
 \label{SErovh}
 \ee
In spite of being manifestly non-Hermitian, the latter model still
predicts a measurable, strictly real and discrete spectrum of
energies which is bounded from below \cite{DDT}.

From our present point of view it is remarkable that whenever one
uses the complexified coordinates (\ref{line}) in
eq.(\ref{SErovh}), the role of the singularity at $r=0$ becomes
virtually irrelevant. The solvable spiked-harmonic-oscillator
illustration of this observation has been described in our letter
\cite{ptho} where the term $\ell(\ell+1)/r^2$ has been noticed
bounded at any nonzero shift $\varepsilon>0$. Still, to our great
surprise, the spectrum of energies of such a spiked harmonic
oscillator, albeit non-equidistant, proved given by a compact
formula  and remained real and discrete.

A step beyond the scope of eq.~(\ref{SErovh}) has been made by
Sinha and Roy \cite{Anjana}. These authors were able to complement
the elementary harmonic-oscillator example $V(r)=r^2+const/r^2$ of
ref. \cite{ptho} with the single centrifugal-like ``spike" at
$r=0$ by a series of its  exactly solvable supersymmetric partners
containing a pair of the two left-right symmetric spikes,
 \ben
 V(r) \sim \frac{g}{(r-c)^2} + \frac{g}{(r+c)^2}  +
 {\rm less\ singular\ terms}
 \een
or a left-right symmetric triplet of the spikes,
 \ben
 V(r) \sim \frac{g'}{(r-c')^2}+ \frac{h}{r^2}  + \frac{g'}{(r+c')^2}  +
 {\rm less\ singular\ terms}
 \een
etc. All of these models were defined along the same complex
straight lines of eq.~(\ref{line}).

\subsection{The birth of quantum  toboggans: Wave functions defined over several Riemann sheets
 \label{qutobo}}

Dorey et al \cite{DDT} found it entirely natural to replace the
regular model (\ref{SErovh}) with $\ell=\omega=0$ by its general
singular version where $\ell$ is real and does not vanish. In this
perspective the interaction term $\ell(\ell+1)/r^2$ {\em ceased}
to be related to the angular momenta, of course (cf. also
\cite{BG} in this respect). In \cite{tob} we were inspired by this
generalization and returned to the study of the role of the
parameter $\ell$. Our analysis started from the observation that
the wave functions are {\em analytic} so that they must have the
following general form near the origin,
 \be
 \psi(r)\sim
 c_1r^{\ell+1} + c_2r^{-\ell}\,,\ \ \ \ \ \ \ \
 |r| \ll 1\,.
 \label{fob}
 \ee
At a generic real $\ell$, the {\em natural} domain of definition
of wave functions {\em must be} the whole logarithmic Riemann
surface ${\cal R}$ \cite{BpT} rather than just one of its sheets.

The core of the message of ref. \cite{tob} was a mere formulation
of a few consequences of the latter facts. We emphasized that
immediately after one allows a complexification of the coordinates
in Schr\"{o}dinger equations, there remains no tenable reason for
the limitation of our attention to the integration paths of the
straight-line form [cf. eq. (\ref{line})] or of the form where a
smooth $x-$dependence of $\varepsilon (x) = \varepsilon (-x)>0$ is
admitted,
 \be
 r = r(x)=x-{\rm i}\varepsilon(x)\,,
 \ \ \ \ \ \
 \ \ \ \ \ x
 \in (-\infty,\infty)\,
 \label{neline}
 \ee
(cf., e.g., ref.~\cite{oparabolas} for illustrative examples).

\begin{figure}[h]                     
\begin{center}                         
\epsfig{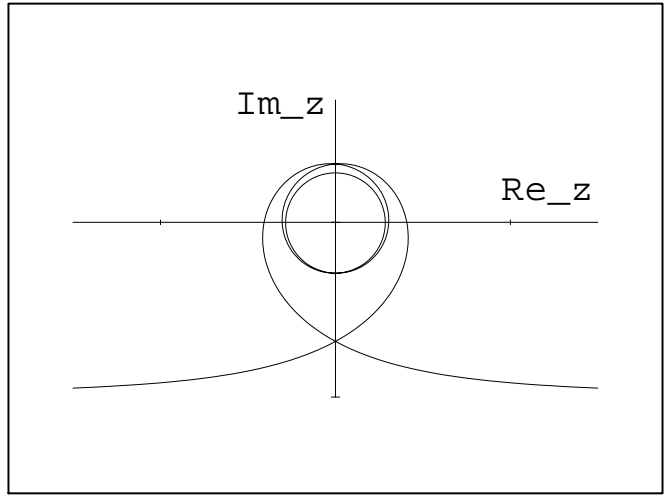}
\end{center}                         
\vspace{-2mm} \caption{Sample tobogganic curve which three times
encircles the origin.
 \label{fione}}
\end{figure}

At this moment, what remained for us to do was an extension of the
oversimplified recipe (\ref{neline}). We started from the
equivalent formula
 \be
 r[x(\gamma)]
 =-{\rm i}\,\varrho(\gamma)\,e^{{\rm i}\gamma}\,,
 \ \ \ \ \
  \ \ \ \ \
 \varrho(\gamma)=\frac{\varepsilon[x(\gamma)]}{\cos \gamma}\,,
 \ \ \ \ \
 \,
 \label{linebb}
 \ee
where $r \in l\!\!C$ denotes the usual complex coordinate (i.e.,
an element of a cut complex plane) and where we merely
reparametrized our original real parameter $x =
x(\gamma)=\varepsilon\,\tan \gamma$ using an  angular variable
$\gamma \in (-\pi/2,\pi/2)$.

In the next (and final) step we allowed the angle to run to
neighboring Riemann sheets. This was achieved by the mere removal
of the restriction on the range of the angle. Alternatively, we
may keep the range of $\gamma \in (-\pi/2,\pi/2)$ and modify just
slightly the definition of the complex, tobogganic curve of the
multisheeted coordinates, yielding, at an arbitrary winding number
$N$,
 \be
 z^{[N]}[x(\gamma)]
 =-{\rm i}\,\varrho^{2N+1}(\gamma)\,e^{{\rm i}(2N+1)\gamma}\,
  \label{linec}
 \ee
This is the formula we need. It specifies a spiral-shaped curve
which may be interpreted as lying on a given logarithmic Riemann
surface ${\cal R}$ which pertains to a given analytic
eigenfunction $\psi(z)$ of our tobogganic Hamiltonian with the
single singularity in the origin (for illustration, we choose
$N=3$ in our illustrative Figure \ref{fione}).

Let us agree that we shall reserve the symbol $z$ for the elements
of ${\cal R}$. Then, in the vicinity of the origin all our wave
functions will have the same generic form (\ref{fob}) and only the
angle in eq.~(\ref{linec}) will mark the distinction between the
separate Riemann sheets. Hence, we can read the pictures of the
curves (\ref{linec}) (cf. Figure \ref{fione}) as encircling  the
essential singularity in the origin at a generic $\ell$. In
effect, these curves extend over an $(N+1)-$plets of the
neighboring Riemann sheets of ${\cal R}$. In the same sense, we
may also write the tobogganic Schr\"{o}dinger equations in the
form
 \be
 \left [-\frac{d^2}{dz^2}
 +\frac{\ell(\ell+1)}{z^2}+\omega^2\,z^2+{\rm
 i}z^3\right ]\,\varphi^{[N]}_n(z)=E^{[N]}_n(\omega)\,\varphi^{[N]}_n(z)
 \,
 \label{SErovhbe}
 \ee
which is formally defined on the Hilbert space
$I\!\!L_2(z^{[N]}[x(-\pi/2,\pi/2)]$ of functions which are
quadratically integrable along the spirals $ z^{[N]}(x)\in {\cal
R}$ which, by assumption, $N-$times encircle the origin.

Whenever we choose a potential $V(z)$ which is analytic at $z\neq
0$ it is not difficult to realize that the existence of the
bound-state solutions $\varphi^{[N]}_n(z)$ of eq.~(\ref{SErovhbe})
will be guaranteed by the following asymptotic boundary
conditions,
 \be
 \varphi^{[N]}_n \{z^{[N]}[x(\pm \pi/2)]\}=0\,.
 \ee
These two constraints must be considered as lying on the different
Riemann sheets. Figure \ref{fione} with $N=3$ may be consulted for
illustrative purposes again.

A purely numerical solution of differential  eq.~(\ref{SErovhbe})
can be constructed by the standard shooting methods. Indeed, one
would have to start at $\gamma \approx -\pi/2$, i.e.,  on the left
asymptotic branch of the curve of Figure \ref{fione} in our
illustration. One has to integrate and move along the curve
$z^{[3]}(x)$ towards the origin. The integration path has to
change, three times, the Riemann sheet before completing the full
triple rotation around the origin and before reaching, finally,
the right asymptotic branch of the integration path at $\gamma
\approx +\pi/2$.

In the next section, a non-numerical reinterpretation of such a
naive recipe will be proposed and advocated.

\section{Rectified tobogganic Schr\"{o}dinger equations in ${\cal H}^{(auxiliary)}$
\label{4.}}

One of the most efficient methods of the numerical solution of the
tobogganic Schr\"{o}dinger equation (\ref{SErovhbe}) may be based
on an auxiliary change of the coordinates complemented by the
parallel modification of the operators and wave functions
\cite{tob}.

In this spirit we are allowed to change the variables from $z \in
{\cal R}$ to $r \in l\!\!C$ and {\it vice versa}. For the sake of
definiteness let us base this change on the conformal mapping
 \be
 {\rm i}r = ({\rm i}z)^\alpha\,,
 \ \ \ \ \ \ \ \ \alpha=\frac{1}{2N+1}\,,
 \ \ \ \ \ N = (0,)\, 1, 2, \ldots
 \label{chov}
 \ee
and let us assume that the ``rectified" coordinate $r$ lies simply
on the line of eq.~(\ref{line}). The related toboggan-shaped
spirals coincide with $z = z^{[N]}(x)$ at a constant $\varepsilon
\neq \varepsilon(x)$. Still, the tobogganic paths resemble the
straight lines at the large real parameters,
 \be
 z^{[N]}(x)=
 -{\rm i}
 \left \{{\rm i}r^{[0]}(x)
 \right \}^{2N+1}= (-1)^N\,x^{2N}\,
 \left [x-(2N+1){\rm i}\varepsilon + {\cal O}\left (\frac{1}{x}
  \right ) \right ]
  \label{patho}
 \,.
 \ee
The latter curves can be interpreted as integration paths of new
type. The resulting quantum system generated by the usual
Hamiltonian but {\em redefined} along the new, topologically
nontrivial paths carries the name of quantum toboggan.

Let us now return to our tobogganic model (\ref{SErovhbe}) with
the cubic interaction and with the first nontrivial choice of the
winding number $N=1$. Via the change of variables (\ref{chov})
this model becomes equivalent to its straight-line rectified
rearrangement
 \ben
 \left [-\frac{d^2}{dx^2}
 +\frac{L(L+1)}{r^2(x)}
  +9\omega^2\,r^{10}(x)
 +9\,{\rm
 i}r^{13}(x)
 \right ]\,\psi_n[r(x)] =
 \ \ \ \ \ \ \ \ \ \ \
 \ \ \ \ \ \ \ \ \ \ \
 \een
 \be
 \ \ \ \ \ \ \ \ \ \ \
 \ \ \ \ \ \ \ \ \ \ \
 \ \ \ \ \ \ \ \ \ \ \
 \ \ \ \ \ \ \ \ \ \ \
 = 9\,E^{[N]}_n(\omega)\,
 r^{4}(x)\,\psi_n[r(x)]
 \,
 \label{SErov}
 \ee
where $r(x)=x-{\rm i}\varepsilon$ and where
 \ben
 \psi_n(r)=z^{-N/(2N+1)}\,
 \varphi_n^{[N]}(z)\,,
 \ \ \ \ \ \ \
 L=({2N+1})\,
 \left (\ell+
 \frac{1}{2}
 \right )-\frac{1}{2}
 \,.
 \een
The real one-dimensional variable $ x \in (-\infty,\infty)$ can be
perceived as a new coordinate. In this way we managed to return
our tobogganic toy model from its fairly exotic representation
(\ref{SErovhbe}) using wave functions $\varphi^{[N]}_n(z)$ defined
on complex spirals $z= z^{[N]}(x)$ to its less exotic though still
non-selfadjoint differential-equation representation (\ref{SErov})
with wave functions $\psi_n[r(x)] $ which lie in the entirely
standard Hilbert space $I\!\!L_2(I\!\!R)\ \equiv\ {\cal
H}^{(auxiliary)}$.

The tobogganic generalization  of the integration paths
(\ref{line}) is unexpectedly nontrivial. Firstly, its consequences
may prove fairly counterintuitive. For example, one encounters
bound states called quantum knots \cite{QK} even when the
potential itself vanishes. Secondly, its mathematical essence
immediately inspires various further generalizations. For example,
one can introduce tobogganic systems containing more than one
branch point \cite{2bp} and/or describing certain less usual forms
of the scattering~\cite{toboscatt}.
%

\subsection{Left and right eigenstates in ${\cal
H}^{(auxiliary)}$\label{4.1}}

Our Schr\"{o}dinger eq.~(\ref{SErov}) is to be solved in the
auxiliary, unphysical but ``user-friendly" Hilbert space ${\cal
H}^{(auxiliary)}$. This illustrates the generic situation in
which, in general, the energy term is absent (i.e., where the
``new energy" is strictly zero). This is one of the most important
immediate consequences of the rectification transformation
(\ref{chov}) applied to the original tobogganic differential
equation [i.e., to eq.~(\ref{SErovhbe}) in our illustrative
example].

In the language of functional analysis, our sample Schr\"{o}dinger
eq.~(\ref{SErov}) still has the form of the ordinary differential
equation,
 \be
 H^{}\,|\,{n}\rangle
 =E_n\,W\,|\,{n}\rangle\,
 \label{sturmed}
  \ee
where the function $W[r(x)]=9\,r^4(x) \,\equiv\,9\,(x-{\rm
i}\varepsilon)^4$ represents an example of the so called
weight-operator in the so called generalized eigenvalue problem
(\ref{sturmed}).

One of the most striking formal features of our tobogganic
Schr\"{o}dinger eq.~(\ref{SErov}) in its version (\ref{sturmed})
obtained by the change of variables can be seen in its manifest
non-Hermiticity in the usual Dirac's sense. More explicitly, in
our Hilbert space ${\cal H}^{(auxiliary)}$ where the Hermitian
conjugation is defined as the usual transposition plus complex
conjugation we have $H \neq H^\dagger$ and $W \neq W^\dagger$ of
course.

This means that whenever we return to the standard Dirac's
conventions we must be much more careful when using the
ket-symbols $|\cdot\kt$ (denoting the elements of
$I\!\!L_2(I\!\!R) \equiv {\cal H}^{(auxiliary)}$) and the
bra-symbols $\br \cdot|$ denoting the corresponding linear
functionals of the same Hilbert space which is, of course,
self-dual.

The first example of the possible ambiguity caused by an
inappropriate notation emerges when we denote the $n-$th
eigenstate of $H$ by the ket $|\,{n}\rangle$ [cf.
eq.~(\ref{sturmed})]. Although the corresponding linear functional
is, of course, the bra $\langle n\,|$, it is {\em not}
proportional to the solution of the Hermitian conjugate version of
the equation. Thus, for the elements of  ${\cal H}^{(auxiliary)}$
which satisfy the latter, ``complementary" Schr\"{o}dinger
equation we {\em definitely have to} introduce {\em another}
symbol.

In what follows we shall accept the conventions advocated in
\cite{SIGMA} and specifying the latter eigenvectors (sought as the
elements of {\em the same} auxiliary Hilbert space ${\cal
H}^{(auxiliary)}$ again) by the mere doubling of the bra or ket
symbols. This leads to the following compact form of the
``independent" generalized eigenvalue problem for our
``independent" operators $H^\dagger$ and  $W^\dagger$,
 \be
 H^{\dagger}\,|\,{n}\rangle\!\rangle
 =E_n^*\,W^{\dagger}\,|\,{n}\rangle\!\rangle\,.
 \label{sturedd}
  \ee
Of course, we may rewrite the latter equation in the more natural,
conjugate form where the operators would act to the left,
 \be
 \langle\!\langle\,n\,|\,H^{}
 =\langle\!\langle\,n\,|\,W^{}\,E_n\,.
 \label{uredd}
  \ee
In the language of mathematics, the latter version of the second,
non-equivalent Schr\"{o}dinger equation is solved directly in the
dual vector space which might be denoted by the symbol $\left
({\cal H}^{(auxiliary)} \right )^\dagger$ if necessary.

Let us now emphasize that in general, the rectification mediated
by an appropriate change of variables in the corresponding
differential Schr\"{o}dinger equation is an invertible
transformation of our original tobogganic eigenvalue problem where
the spectrum of the energies $E^{[N]}_n$ has been expected real
and where the Hamiltonian and its conjugate operator have been
assumed isospectral. These questions have already thoroughly been
discussed elsewhere \cite{tob}. In the present context this only
means that all the key spectral properties of our toboggans will
be inherited by the conjugate pair of the {\em rectified}
eigenvalue problems (\ref{sturmed}) and (\ref{sturedd}) or
(\ref{uredd}) in ${\cal H}^{(auxiliary)}$ (cf. also a few relevant
remarks added in Appendix C).

A marginal remark may be added concerning the rectified toboggans
in a purely pragmatic numerical context where our coefficients
$E^{[N]}_n$ need not necessarily be called the energies and where
our ket eigenstates can equally well be perceived as  the ``alias"
zero-energy bound states generated by the Hamiltonian
$H^{(alias)}=H-E^{[N]}_nW$. In this sense, for example, our
particular illustrative eq.~(\ref{SErov}) would acquire the
standard zero-energy form where one searches for certain
particular coupling strengths at the quartic component of the
potential.

We may summarize that our task degenerates now to the quantization
of the real coefficients $E^{[N]}_n$  at the weight factor $W$ of
a well-specified toboggan-oscillator physical origin. We may be
sure that the nontriviality of $W \neq I$ cannot become a source
of any unexpected mathematical difficulties precisely due to the
existence of the possibility of an invertibility of our
rectification transformation. Of course, {\em without} such a
pull-back to the original analytic equation for a toboggan the
study of our equations containing some more general choices of $W$
would open a Pandora's box of many interesting as well as fairly
difficult mathematical questions. Here, we need not and will not
extend the discussion of these questions beyond our present very
specific framework of tobogganic models.

\subsection{Biorthogonality and completeness  in ${\cal
H}^{(auxiliary)}$\label{biolo} }

Let us now re-number the states in  ${\cal H}^{(auxiliary)}$
directly by their energies, $|\,{n}\rangle = |\,E_{n}\rangle =
|\,\lambda \rangle $ with the new set of the real indices $
\lambda \in \{ E_{0},\,{E_1},\,\ldots\}$. In such a reduced
notation our two Schr\"{o}dinger equations (\ref{sturmed}) and
(\ref{sturedd}) will read
 \be
 H^{}\,|\,{\lambda}\rangle
 =\lambda\,W\,|\,{\lambda}\rangle\,
 \label{sturmjed}
 \,,\ \ \ \ \ \ \ \ \
 \langle\!\langle\,\lambda'\,|\,H^{}
 =\langle\!\langle\,\lambda'\,|\,W^{}\,\lambda'\,
 \label{sturmdva}
 \ee
Under certain general assumptions the left and right generalized
eigenstates of our $H$ may be shown to be mutually biorthogonal
with respect to the weight-operator $W$,
 \be
  \langle\!\langle\,\lambda'\,|\,W\,|\,{\lambda}\rangle
 =\delta_{\lambda,\lambda'}\cdot\,\sigma_\lambda\,,
 \ \ \ \ \ \ \
 \sigma_\lambda= \langle\!\langle\,\lambda\,|\,W\,|\,
 {\lambda}\rangle\,.
 \label{bior}
 \ee
In parallel, the completeness relations in ${\cal
H}^{(auxiliary)}$ will be assumed in the form
 \be
 I = \sum_{\lambda}\,|\,\lambda\,\kt \,
 \frac{1}{
 \sigma_\lambda}
 \,
  \langle\!\langle\,\lambda\,|\,
 W\,.
 \label{bico}
 \ee
Technical details of these formulae have been thoroughly discussed
in ref.~\cite{SIGMA}. Their essence can briefly be summarized as
reflecting the fact that {\em both} the independent
Schr\"{o}dinger equations in (\ref{sturmjed}) are linear and
homogeneous. As a consequence, any ``initial", fixed set of these
solutions marked by an auxiliary superscript $^{[1]}$ can be
replaced by another set containing just some {\em different}
multiplication constants and marked by some other superscript.

In what follows, we shall employ the notation convention of
ref.~\cite{SIGMA} where the initial set of eigenvectors
$|\,\lambda\,\kt =|\,\lambda\,\kt^{[1]}$ and
$\langle\!\langle\,\lambda'\,|=
\,^{[1]}\!\langle\!\langle\,\lambda'\,|$ is allowed to be
generalized in accordance with the most elementary rule
 \be
  |\,\lambda\,\kt =|\,\lambda\,\kt^{[\vec{\kappa}]}
  =|\,\lambda\,\kt^{[1]}/\kappa_\lambda\,,
 \ \ \ \ \ \langle\!\langle\,\lambda'\,|=
 \,^{[\vec{\kappa}]}\!\langle\!\langle
 \,\lambda\,|
 =\kappa_{\lambda}\,\,^{[1]}\!\langle\!\langle\,\lambda\,|\,
 \label{iruna}
 \ee
where $ \lambda= E_{0},\,{E_1},\,\ldots$ and where any element of
the sequence of (complex) constants
$\vec{\kappa}=\{\kappa_{\lambda_0},
\,\kappa_{\lambda_1},\,\ldots\}$ is {\em arbitrarily variable}.

In a sharp contrast to the Hermitian case where the simple and
double kets coincide, our choice of any numerical factor
$\kappa_{\lambda_j}$ is now virtually unrestricted. Indeed, the
change of any $\kappa_{\lambda_j}$ changes {\em neither} the
biorthogonality relations (\ref{bior}) {\em nor} the completeness
relations (\ref{bico}) and neither the values of coefficients
$\sigma_\lambda$ nor the spectral decomposition of our Hamiltonian
operator,
 \be
 H = \sum_{\lambda}\,W\,|\,\lambda\,\kt
 \,
 \frac{\lambda}{\sigma_\lambda}
 \,\langle\!\langle\,\lambda\,|\,
 W\,.
 \label{bisp}
 \ee
This means that the specific renormalization freedom and the
presence and the free variability of the superscript
$^{[\vec{\kappa}]}$ should always (at least, tacitly) be kept in
mind in what follows.

\section{Formula for the metric in ${\cal H}^{(physical)}$\label{manolo} }

%

\subsection{An update of the metric \label{manoloja}
$\Theta_{(Dirac)}^{(auxiliary)} \, \to \,
\Theta_{(non-Dirac)}^{(physical)}$ at $W \neq I$.}

Due to the relations $H\neq H^\dagger$ and $W \neq W^\dagger$ the
space ${\cal H}^{(auxiliary)}$  {\em cannot } be interpreted and
accepted as the Hilbert space of states of our quantum system.
Instead, we have to follow and modify the recipe which has been
used in the special case where $W=I$ and $H \neq H^\dagger$ in
${\cal H}^{(auxiliary)}$ \cite{Geyer}. Thus, we shall assume that
the correct physical Hilbert space ${\cal H}^{(physical)}$ must be
introduced via a non-unitary though still invertible mapping of
the kets $|\psi\kt \in {\cal H}^{(auxiliary)}$ on their
``spiked-ket" images $|\,\psi \pkt \ \in {\cal H}^{(physical)}$.
In this notation the mapping
 \ben
 \Omega:
 \ {\cal H}^{(auxiliary)}
 \,\longrightarrow\,
 {\cal H}^{(physical)}\,
 \een
will be realized by the operators $\Omega$ which can be written in
the form of the series
 \be
 \Omega=\Omega^{[\vec{\kappa}]}=
 \sum_\lambda
 |\,\lambda \pkt\,\frac{1}{\sigma_\lambda}\
 ^{[\vec{\kappa}]}\!\langle\!\langle\,\lambda\,|\,W\,
 \label{nee}
 \ee
containing not yet normalized (i.e., formally,
superscript-independent) spiked kets $|\,\lambda \pkt$. Under this
convention, the explicit presence of the superscript in
$\Omega^{[\vec{\kappa}]}$ is essential because of the absence of
the cancellation between the numerators and denominators,
 \be
 \Omega^{[\vec{\kappa}]}=
 \sum_\lambda\,
 |\,\lambda \pkt\,\left (\frac{\kappa_{\lambda}}{\sigma_{\lambda}}
 \right )\
 ^{[1]}\!\langle\!\langle\,\lambda\,|\,W\,.
 \label{denomi}
 \ee
This indicates that $\Omega^{[\vec{\kappa}]} \neq
\Omega^{[\vec{\kappa}']}$ for $\vec{\kappa} \neq \vec{\kappa}'$
and that there exists a ``hidden" ambiguity in the definition
 \be
 |\,\lambda\,\pkt = \Omega\,|\,\lambda\,\kt\,\in\,
 {\cal H}^{(physical)}\,,
 \ \ \ \ \
 \pbr\,\lambda\,| = \langle\,\lambda\,|\,
 \Omega^\dagger\,\in\,
 \left ({\cal H}^{(physical)}\right )^\dagger\,
 \label{carymary}
 \ee
of the same eigenstates (or of their arbitrary linear
superpositions) {in their two different representations}.

In ${\cal H}^{(physical)}$ we may now parallel the recipe of
ref.~\cite{Geyer} and introduce the operator $h =
\Omega\,H\,\Omega^{-1}$ which represents our original
non-Hermitian upper-case Hamiltonian $H\neq H^\dagger$ in the new
space. The same type of transformation must be used to generate
also the physical partner $w = \Omega\,W\,\Omega^{-1}$ of our
original non-Hermitian weight operator $W\neq W^\dagger$. These
two lower-case operators are both assumed acting in the physical
space so that they must both be self-adjoint in ${\cal
H}^{(physical)}$,
 \ben
 h^\dagger = \left (\Omega^{-1}
 \right )^\dagger\,H^\dagger\,\Omega^\dagger= h\,,
 \ \ \ \ \ \ \ \ \ \ \
  w^\dagger = \left (\Omega^{-1}
 \right )^\dagger\,W^\dagger\,\Omega^\dagger= w\,.
 \een
This implies  that in ${\cal H}^{(auxiliary)}$ we must have
 \be
 H^\dagger = \Theta\,H\,\Theta^{-1}\,,
 \ \ \ \ \
 W^\dagger = \Theta\,W\,\Theta^{-1}\,
 \label{qh}
 \ee
where we abbreviated $\Theta = \Omega^\dagger\,\Omega$. In the
special case of $W=I$ this conclusion degenerates to the one
presented in \cite{Geyer}.

Let us now insert expressions (\ref{qh}) in our ``second"
Schr\"{o}dinger eq.~(\ref{sturedd}) and compare it with its
``first" form (\ref{sturmed}). Under the standard non-degeneracy
assumption this implies the following elementary proportionality
between the eigenkets of $H^\dagger$ (= eigen-double-kets) and the
metric-premultiplied eigenkets  of $H$ (= eigen-single-kets), both
being, optionally, superscripted in accordance with
eq.~(\ref{iruna}),
 \be
 |\,\lambda\,\kkt^{[\vec{\kappa}]}
  = \Theta^{[\vec{\kappa}]}\,|\,
  \lambda\,\kt^{[\vec{\kappa}]}
  \cdot q(\lambda) \,.
  \label{recall}
  \ee
The explicit knowledge of the metric would be needed to extract
the values of the proportionality constants $q(\lambda)$ here.

\subsection{An update of the eigenstates
 at $W \neq I$\label{4.2}}

After the change $\Omega$ of the Hilbert space {\em both} of
eqs.~(\ref{sturmjed}) degenerate to the {\em same} equation which
is self-adjoint,
 \be
 h^{}\,|\,{\lambda}\pkt
 =\lambda\,w\,|\,{\lambda}\pkt\,.
 \label{lcsturmjed}
 \ee
Under some very general and more or less usual mathematical
assumptions its form enables us to deduce the orthogonality
relations
 \be
 \pbr \lambda\,|\,w\,|\,\lambda'\pkt = \tilde{\sigma}_\lambda
 \,\cdot\
 \delta_{\lambda,\lambda'}
 \,,
 \ \ \ \ \ \ \
 \tilde{\sigma}_\lambda=\
 \pbr \lambda\,|\,w\,|\,\lambda\pkt
 \,.
 \label{biorto}
 \ee
In ${\cal H}^{(physical)}$ the  completeness relations are also
valid,
 \be
 I=
 I^{(physical)} = \sum_{\lambda}\,|\,\lambda\pkt
 \,
 \frac{1}{\tilde{\sigma}_\lambda}
 \,\pbr \lambda\,|\,w
 \label{bicomp}
 \ee
and the lower-case Hamiltonian $h$ can equally easily be assigned
the usual spectral expansion,
 \be
 h = \sum_{\lambda}\,w\,|\,\lambda\pkt\,
 \frac{\lambda}{\tilde{\sigma}_\lambda}
 \,\pbr \lambda\,|\,w\,.
 \label{bispec}
 \label{spre}
 \ee
It is important to add that in the light of our illustrative
example (\ref{SErov}) the spectrum of the operator $w$ (which is,
by construction, isospectral to $W$) will be assumed non-negative.
This will allow us to assume also the positivity of the
self-overlaps $ \pbr \lambda\,|\,w\,|\,\lambda'\pkt $ in
eq.~(\ref{biorto}) etc. Both these assumptions will, of course,
significantly simplify our forthcoming considerations.

All the relations (\ref{biorto}) -- (\ref{bispec}) are of a rather
academic value since all the mathematical manipulations should
preferably be performed in the simpler Hilbert space ${\cal
H}^{(auxiliary)}$, anyhow. Still, all of them improve our insight
in the possible physics behind our models which can solely be
discussed inside the physical Hilbert space ${\cal
H}^{(physical)}$.

The mapping $\Omega$ can be read as returning us back from the
``correct" ${\cal H}^{(physical)}$ to the ``simpler" Hilbert space
${\cal H}^{(auxiliary)}$. In particular, in the light of
eq.~(\ref{carymary}) we may complement the proportionality
relation (\ref{recall})  by another formula,
 \ben
 \br\ \lambda\,|\,\Theta\,=\
 \pbr\,\lambda\,|\,\Omega\,
 \een
which enables us to transfer eqs.~(\ref{biorto})
and~(\ref{bicomp}) to ${\cal H}^{(auxiliary)}$,
 \be
  \pbr \lambda\,|\,w\,|\,\lambda'\pkt \ =
 \langle\,\lambda\,|\,\Theta\,W
 \,|\,\lambda'\,\kt =
 \br\ \lambda\,|\,\Theta\,|\,W
 \,|\,\lambda'\,\kt =
 \tilde{\sigma}_{\lambda}\,\cdot\,
 \delta_{\lambda,\lambda'}\,,
 \label{30}
 \ee
 \be
 I =
 I^{(auxiliary)}= \sum_{\lambda}\,|\,\lambda\,\kt \,
 \frac{1}{
 \tilde{\sigma}_{\lambda}}
 \,
 \br\ \lambda\,|\,\Theta\,
 W\,.
 \ee
A similar translation applies also to the alternative spectral
decomposition
 \be
 H = \sum_{\lambda}\,W\,|\,\lambda\,\kt
 \,
 \frac{\lambda}{
 \tilde{\sigma}_\lambda}
 \,\br\ \lambda\,|\,\Theta\,
  W\,
 \ee
of our Hamiltonians in ${\cal H}^{(auxiliary)}$. In the light of
eq.~(\ref{recall}), the only difference from the respective
eqs.~(\ref{bior}), (\ref{bico}) and~(\ref{bisp}) degenerates to
the following relation between the tilded and untilded overlaps
 \be
 \tilde{\sigma}_\lambda=
 \,\br\ \lambda\,|\,\Theta\,W
  \,|\,\lambda\,\kt
 =\
  \left [\frac{1}{ q(\lambda)} \right ]^*\,
  \sigma_\lambda\,,
  \ \ \ \ \ \ \sigma_\lambda=
  \, \bbr \lambda\,|\,W
  \,|\,\lambda\,\kt\,
  \label{onyx}
  \ee
which are all, incidentally, $\vec{\kappa}-$independent.

\subsection{The ultimate double-series update
 of $\Theta$ at $W \neq I$ \label{quito}  }

Let us return to the simpler Hilbert space ${\cal
H}^{(auxiliary)}$ and, via a suitable multiplier in the
eigen-doublekets $|\,\lambda\kkt$, let us postulate that
$\sigma_\lambda=1$. In parallel, in ${\cal H}^{(physical)}$, the
maximum of simplicity will be achieved by setting
$\tilde{\sigma}_\lambda= 1$ via a premultiplication of our basis
vectors $|\,\lambda\,\pkt\ $ by some suitable numerical constants
at each energy $\lambda$. Thus, we shall have $q(\lambda)=1$ in
eq.~(\ref{recall}) as well as in eq.~(\ref{onyx}), i.e.,
 \be
 \  \pbr \lambda\,|\,w\,|\,\lambda\pkt \ = 1\,.
 \label{37be}
 \ee
Under these updated conventions we shall be allowed to work,
without any loss of generality, with the simplified
$\sigma_\lambda=1$ versions of eqs.~(\ref{bior}) and~(\ref{bico}) 
and with the similarly  simplified representation~(\ref{bisp}) of
the Hamiltonian $H$.

Even though we have no direct access to the metric $\Theta$ and/or
to the overlaps and matrix elements defined in terms of the
``inaccessible" and ``prohibitively complicated" vectors
$|\,\lambda\,\pkt\ \in {\cal H}^{(physical)}$ we can still recall
the definition of the mappings $\Omega$ and write
 \be
 \Theta=\Omega^\dagger\,\Omega=
 \sum_{\lambda,\lambda'}\,
 W^\dagger\,|\,\lambda \kkt\,
 M_{\lambda,\lambda'}\,\bbr\,\lambda'\,|\,W
 \,,
 \ \ \ \ \ \ \ M_{\lambda,\lambda'} =\
\pbr\!\lambda\,|\,\lambda'\!\pkt\
 \,.
 \label{reforum}
 \ee
The not yet known matrix  $M$ of coefficients cannot vary with the
(tacitly present) superscripts $^{[\vec{\kappa}]}$ so that, in the
light of eq.~(\ref{denomi}), there is no cancellation between
numerators and denominators and
 \ben
 \Theta^{[\vec{\kappa}]} \neq
 \Theta^{[\vec{\kappa}']}\ \ \ \ {\rm for}\ \ \ \
 \vec{\kappa} \neq \vec{\kappa}'\,.
 \een
The change of the superscripts will change the metric only via the
eigenvectors of $H^\dagger$ [cf. eq.~(\ref{iruna})] so that we can
rewrite the superscript-dependence of the metric in
eq.~(\ref{reforum}) in the following explicit form
 \be
 \Theta^{[\vec{\kappa}]} =
 \sum_{\lambda,\lambda'}\,
 W^\dagger\,|\,\lambda \kkt^{[1]}
 \,\kappa^*_\lambda\,
 M_{\lambda,\lambda'}\,\kappa_{\lambda'}\  ^{[1]}\bbr\,\lambda'\,|\,W\,.
 \label{repoforum}
 \ee
Now we return to the factorization $\Theta=\Omega^\dagger\Omega$
of the metric and  to identity (\ref{37be}) and definition
(\ref{nee}). We insert all of these formulae in eq.~(\ref{30}).
This gives the relation
 \be
 \langle\,\lambda\,|\,\Theta\,W
 \,|\,\lambda'\,\kt =
 \sum_{\lambda''}\ \pbr \lambda\,|\,\lambda''\pkt\
 \bbr \,\lambda''\,|\,W^2\,|\,\lambda' \kt
 =
 \delta_{\lambda,\lambda'}\,.
  \label{rerelye}
 \label{37bece}
 \ee
Once we succeed in evaluating all the necessary ``input" matrix
elements $\bbr \,\lambda\,|\,W^2\,|\,\lambda' \kt$, the ``missing"
matrix of coefficients $M_{\lambda,\lambda'}=\, \pbr
\lambda\,|\,\lambda'\pkt\,$ will be defined as its inverse,
 \be
 M = S^{-1}\,, \ \ \ \ \ \ S_{\lambda,\lambda'}=\bbr
\,\lambda\,|\,W^2\,|\,\lambda' \kt\,.
 \ee
This result is to be inserted in formula (\ref{repoforum}) for the
metric $\Theta$. Our task is completed.

\section{Summary \label{sum} }

Many quantum Hamiltonians with real spectra which appeared
manifestly non-Hermitian in the current Dirac's sense were
recently re-assigned a new, consistent probabilistic
interpretation mediated by a new metric $\Theta$ in the physical
Hilbert space of states. For certain models of this type
(exemplified here by ``quantum toboggans"), the Mostafazadeh's
spectral-expansion formula for $\Theta$ \cite{Ali1} ceases to be
applicable because their Schr\"{o}dinger equation acquires the
generalized eigenvalue-problem form $H\psi = EW\psi$ containing an
invertible though not necessarily positive-definite weight $W\neq
I$. For all of these models we derived the necessary generalized
spectral-expansion formula for $\Theta$.

We started our considerations from a given non-Hermitian
tobogganic Hamiltonian playing the usual role of the generator of
the time evolution but acting, very unusually, along a certain
topologically nontrivial path $z^{[N]}$ of complex coordinates.
Via a suitable change of variables we achieved a rectification of
this path and obtained a much more usual non-Hermitian
representation $H$ of the Hamiltonian operator in an auxiliary
Hilbert space $I\!\!L_2(I\!\!R)$.

For the purely pragmatic reasons we constrained our attention to
the mere ``first nontrivial" toboggans with the single branch
point. Our formulation of the corresponding bound-state problem
has been facilitated by their rectification. Still, as long as the
change of the variables induced a nontrivial weight operator $W$
in our Schr\"{o}diger equation, the standard recipes of dealing
with similar situations proved inapplicable and we were forced to
modify them accordingly. Fortunately, via a subsequent non-unitary
mapping $\Omega$ we were able to replace the non-Hermitian ``upper
case" operators $H$ and $W$ by their respective ``lower-case"
avatars $h$ and $w$ defined as Hermitian in another, physical
Hilbert space of states ${\cal H}^{(physical)}$.

We should emphasize that in similar models, the non-unitary
correspondence between two Hilbert spaces is of a key importance.
Its main purpose lies in a {\em decisive simplification} of
mathematics (i.e., e.g., of the solution of equations) in one of
the spaces, combined with a facilitated return to the consistent
probabilistic interpretation of the system in the other one. In
this sense, our tobogganic models also fit very well the basic
methodical premise that the lower-case, ``correct" representation
$h$ of the Hamiltonian appears,  in the purely technical terms,
{\em too complicated} in comparison with $H$.

Having paid our main attention to the correct physical
interpretation of the elementary though nontrivial tobogganic
models we formulated a straightforward $W\neq I$ generalization of
the known $W=I$ theory of ref.~\cite{Geyer}. We succeeded in a
derivation of the explicit formula for the metric operator
$\Theta$. We revealed that the coefficients in this formula
coincide with an inverse of certain matrix $S$ representing the
square of the weight operator $W$ in a certain basis. In this
sense, our formula degenerates to the older theories in the limit
$W\to I$ where $S$ becomes a diagonal matrix.

Another conclusion resulting from our formula for $\Theta$ is that
without essential changes, the ambiguity problem in the assignment
of $\Theta$ to a given set of observables $A_j \neq A_j^\dagger$
(containing the Hamiltonian $H=A_0$, etc) survives the transition
to the systems with a nontrivial weight operator $W$. Indeed, when
we only take the Hamiltonian $H=A_0$ into consideration, the same
infinite sequence of arbitrary complex parameters $\kappa_\lambda$
enters the formula for the metric [cf. eq.~(\ref{repoforum})] in
{\em both} the $W=I$ and $W \neq I$ scenarios.

\subsection*{Acknowledgement}

Work supported by GA\v{C}R, grant Nr. 202/07/1307, Institutional
Research Plan AV0Z10480505 and by the M\v{S}MT ``Doppler
Institute" project Nr. LC06002.

\newpage

\section*{Appendix A: Non-unitary maps $\Omega$ in Quantum Mechanics
\label{rvnonu} }

The most disturbing feature of all the Dyson-like invertible
mappings $\Omega$ of spaces as studied in ref. \cite{Geyer} is
their non-unitarity, $\Omega^\dagger \neq \Omega^{-1}$. Due to it,
the ``tractable" Hamiltonians $H$ in the space ${\cal
H}^{(auxiliary)}$ are manifestly non-Hermitian. This is not too
essential of course -- for all the physical predictions one can
always return to the (in our present notation, lower-case)
pull-backs of $H$s in ${\cal H}^{(physical)}$,
 \be
 h = \Omega\,H\,\Omega^{-1}\,.
 \label{pullback}
 \ee
Even if the operators (\ref{pullback}) themselves remain, by
assumption, too complicated for computational purposes, they are
still observable, i.e., self-adjoint in their own Hilbert space
${\cal H}^{(physical)}$,
 \ben
 h^\dagger = \left (\Omega^{-1}
 \right )^\dagger\,H^\dagger\,\Omega^\dagger= h\,.
 \een
From this relation one deduces that
 \be
 H^\dagger = \Theta\,H\,\Theta^{-1}\,,
 \ \ \ \ \ \ \ \ \ \Theta = \Omega^\dagger\,\Omega\,.
 \label{qh1}
 \ee
We should note that our use of the symbol $\Theta$ for the
``metric operator" is not too widespread, being equivalent to $T$
or $\eta_+$ or ${\cal CP}$ or $e^Q$ of refs.~\cite{Geyer} or
\cite{Ali1} or \cite{BB2} or \cite{Jones}, respectively.

The identification of all the other observables  $A_1, A_2,
\ldots$ in the ``tractable" Hilbert space ${\cal H}^{(auxiliary)}$
remains entirely straightforward. Once we return to the derivation
of eq.~(\ref{qh1}) we immediately see that all of these operators
must obey the same intertwining rule as the Hamiltonian,
 \be
 A_j^\dagger \Theta = \Theta \,A_j\,.
 \ee
In order to avoid confusion or lengthy explanations (``in which
space?"), the authors of ref.~\cite{Geyer} suggested to call all
these ``admissible" operators of observables ``quasi-Hermitian".

\section*{Appendix B: The birth of models with complex coordinates}

It is well known that in nuclear, atomic and molecular systems
many features of the observed bound-state spectra can often be
very well understood and explained via an elementary differential
Hamiltonian
 \be
 H^{(rad)}= -\frac{d^2}{dr^2}
 +\frac{\ell^{(rad)}(\ell^{(rad)}+1)}{r^2}+V^{(rad)}(r)
 \label{SEr}
 \ee
(units $\hbar=2m=1$) where the phenomenological requirements are
usually reflected by an appropriate adaptation of the real
potential $V^{(rad)}(r)$ defined along the real and nonnegative
coordinate $r \in (0,\infty)$. For pragmatic reasons this
potential is often chosen as confining. Then, the spectrum itself
remains always ``acceptable", i.e., real and discrete and bounded
below.

It is rather surprising to reveal that the spectrum {\em can} stay
real, discrete and bounded below even for certain {\em
complexified} potentials and/or paths of coordinates. For
illustration one might recollect the early works by Caliceti et al
\cite{Caliceti} or by Buslaev and Grecchi \cite{BG}. Before all of
these apparent ``curiosities" fell into oblivion there appeared,
in 1998, the influential letter by Bender and Boettcher \cite{BB}
which we already cited above. This letter offered a numerically
and semi-classically inspired hypothesis that the energies may be
expected to stay real and discrete for {\em many} complex
potentials. An unexpectedly intensive growth of interest in the
similar models followed (cf. \cite{Carl} or \cite{Shin} for more
references).

In the context of the similar concrete examples it soon appeared
to be easy to prove the reality of the energies for a large family
of the exactly solvable complex potentials \cite{Geza}.
Subsequently, in 2001, Dorey et al \cite{DDT} rigorously proved
also the Bender's and Boettcher's hypothesis for the original
class of the field-theory-related power-law complex potentials of
ref.~\cite{BB}. They were even able to introduce more parameters
(generalizing also the (half-)integer $\ell^{(rad)}$ in
eq.~(\ref{SEr}) \cite{BG} to any real parameter $\ell$) and
succeeded in an {\em explicit} specification of the boundaries of
the domains ${\cal D}$ of these parameters where the reality of
the whole spectrum is guaranteed.

\section*{Appendix C: The role of ${\cal PT}-$symmetry
     \label{derte}}

An important formal feature of our illustrative tobogganic
examples can be seen in the so called ${\cal PT}-$symmetry (cf.
\cite{Carl}) where ${\cal P}$ denotes space-reversal (i.e.,
parity) while ${\cal T}$ represents the time reversal (i.e., in
effect, Hermitian-conjugation antilinear operator). Such a feature
of the toboggan-like models can be interpreted here as the {\em
doublet} of the parity-pseudo-Hermiticity properties
 \be
 H^\dagger = {\cal P}\,H\,{\cal P}^{-1}\,,
 \ \ \ \ \ \ \
 W^\dagger = {\cal P}\,W\,{\cal P}^{-1}\,.
 \label{gepe}
 \ee
The main role of this ``generalized symmetry" lies in possible
simplification of the necessary proof that all the energies remain
real, $E_n=E^*_n$, i.e., that the states of the underlying quantum
system remain observable.

In addition, the generalized ${\cal PT}-$symmetry (\ref{gepe})
will enable us to replace eq.~(\ref{sturedd}) by its equivalent
representation
 \be
 H\,{\cal P}^{-1}\,|\,{n}\rangle\!\rangle
 =E_n\,W\,{\cal P}^{-1}\,|\,{n}\rangle\!\rangle\,.
 \label{assturedd}
  \ee
From eq.~(\ref{sturmed}) and from another simplifying assumption
that the spectrum is nondegenerate we immediately deduce that we
must have
 \be
 |\,{n}\rangle\!\rangle=
 {\cal P}^{}\,|\,{n}\rangle\,Q_n
 \label{asstur}
  \ee
where the coefficients of proportionality $Q_n$ called
quasi-parities \cite{pseudo} are, in principle, arbitrary. Still,
one must keep in mind that once we postulate the standard
biorthogonality and completeness relations,
 $
 \bbr n'\,|\,n\kt = \delta_{nn'}\,,
 $ and $
 I = \sum_n\ |\,n\kt\,\bbr\,n\,|
 $,
we are {forced to define} the quasiparities $Q_n=Q_n(\kappa_n)$,
at all the energy levels, {in terms of the matrix elements of the
parity operator},
 \ben
 Q_n(\kappa_n)=\frac{1}{^{[\kappa_n]}\br
 n\,|\,{\cal P}\,|\,n\kt ^{[\kappa_n]}}\,.
 \een
Thus, the knowledge of these matrix elements fixes the values of
quasiparities while equation~(\ref{asstur}) replaces the second
Schr\"{o}dinger eq.~(\ref{sturedd}) as it becomes an {\em explicit
definition} of its solutions. This fact can, of course, shorten
the ultimate construction of $\Theta$ very significantly.

For illustrative purposes, non-Hermitian anharmonic oscillators
are often being chosen in the literature. As we already mentioned,
the birth of interest in these models  dates back to the letter
\cite{BB}. In this letter, one of the key ideas of all the
subsequent developments of the subject has been presented via the
most elementary harmonic oscillator Hamiltonian $H = p^2+x^2$
defined in the usual Hilbert space $I\!\!L_2(I\!\!R)$. It has been
noticed there that this Hamiltonian has a real spectrum
($E_n=2n+1$) and that it is, at the same time, ${\cal
PT}-$symmetric (for the time being, we can simply understand the
latter concept as a left-right invariance of $H$ in the complex
plane of $x$). In the next step Bender and Boettcher ({\it loc.
cit.}) emphasized that both the latter features remain unchanged
when one adds a manifestly non-Hermitian linear interaction term
${\rm i}x$. In the third step they re-analyzed this type of
correspondence on several other examples and conjectured that
there may exist {\em many} complex potentials for which the
reality of the spectrum can be ``deduced" from the ${\cal
PT}-$symmetry of the system

An important source of the appeal of using the phenomenological
operators of the latter class should be sought in the feasibility
of working with them in the ``unphysical" Hilbert space
$I\!\!L_2(I\!\!R)\,\equiv\,{\cal H}^{(auxiliary)}$ where they
remain ``sufficiently elementary" (i.e., typically, ordinary
differential) operators. In the context of field theory, this idea
of simplicity motivated, {\it inter alia}, the study of
Hamiltonian densities containing the most elementary
parity-violating interaction terms, typically of the form ${\rm
i}g\varphi^3(\vec{x},t)$ \cite{BM}. In the context of mathematics,
many properties of these models found their explanation by means
of a return to their most elementary but already nontrivail
quantum-mechanical predecessors. For example, in the late eighties
the one-dimensional non-Hermitian potential
 \be
 V^{(CGM)}(r)=\omega^2\,r^2+{\rm i}r^3\,
 \label{Cali}
 \ee
was chosen for detailed perturbation studies
\cite{Caliceti,Alvarez}.

\end{document}